%% file: main.tex
\documentclass{article}

\PassOptionsToPackage{numbers, compress}{natbib}

\usepackage[creativeai, final]{neurips_2025}

\usepackage[utf8]{inputenc} 
\usepackage[T1]{fontenc}    
\usepackage{hyperref}       
\usepackage{url}            
\usepackage{booktabs}       
\usepackage{amsfonts}       
\usepackage{nicefrac}       
\usepackage{microtype}      
\usepackage{xcolor}         
\usepackage{adjustbox}
\usepackage{algorithm}
\usepackage{algpseudocode}
\usepackage{amsmath}
\usepackage[capitalize,noabbrev]{cleveref}
\usepackage{caption}
\usepackage{subcaption}     

\title{Stroke Patches:\\Customizable Artistic Image Styling Using Regression}

\author{%
  Ian Jaffray \\
  \texttt{ian.jaffray@outlook.com} \\
  \And
  John Bronskill \\
  University of Cambridge \\
  \texttt{jfb54@cam.ac.uk} \\
}

\begin{document}

\maketitle

\begin{abstract}
We present a novel, regression-based method for artistically styling images.
Unlike recent neural style transfer or diffusion-based approaches, our method allows for explicit control over the stroke composition and level of detail in the rendered image through the use of an extensible set of \textit{stroke patches}.
The stroke patch sets are procedurally generated by small programs that control the shape, size, orientation, density, color, and noise level of the strokes in the individual patches.
Once trained on a set of stroke patches, a U-Net based regression model can render any input image in a variety of distinct, evocative and customizable styles. 
\end{abstract}

\section{Introduction}
\label{sec:introduction}
Transforming images so that they appear to be rendered in an artistic style has a long history.
Gallery Effects \citep{plugin,jaffray1991apparatusa,jaffray1991apparatusb,jaffray1991apparatusc} employed a graph of image processing operators to render images as watercolor, charcoal, and pen and ink, among others.
\citet{haeberli1990paint} created impressionist renderings by specifying the position, color, size, direction, and shape of each stroke driven by features of the source image.
\citet{gatys2015neural} introduced the concept of neural style transfer by using neural networks to separate the content and style of an image, allowing the style from a source image to be imparted on a target image while maintaining the content.
%
%
More recently, diffusion based methods \citep{pmlr-v37-sohl-dickstein15,ho2020denoising} and their implementation in various products \citep{podell2023sdxl,achiam2023gpt,Google2025Gemini} enables the styling of an input image by describing the desired artistic appearance using a natural language prompt.

While the recent machine learning-based approaches excel at capturing the overall \textit{feel} of a specified style, they lack control over the composition of the strokes and the amount of detail in the rendered images.
We propose a complementary approach to artistically styling images that allows explicit and extensible control over image detail through the use of procedurally generated \textit{stroke patches}.
By controlling the shape, size, orientation, density, color, and noise level of the strokes in the stroke patches, a wide variety of expressive artistic looks can be achieved.

Our contributions are twofold.
The first is the concept of a customizable, procedurally generated stroke patch, which defines an artistic look and the level of detail in the stylized output.
The second is using a regression-based approach that, once trained on a set of stroke patches, learns to map continuous tone regions of an image to the discrete strokes specified in the stroke patches.

\cref{sec:method} defines a stroke patch and how to create them, the model architecture, and the details of training and inference.
\cref{sec:results} depicts a range of outputs from the method and their associated stroke patches, as well as the effect of varying certain stroke patch parameters.
Finally, \cref{sec:discussion} summarizes our work and places it in perspective with existing artistic image rendering methodologies.
\section{Method}
\label{sec:method}
In this section, we define the notion of a set of stroke patches, the overall system architecture, and detail training and inference.
\begin{figure}[th]
\begin{center}
\includegraphics[width=0.8\textwidth]{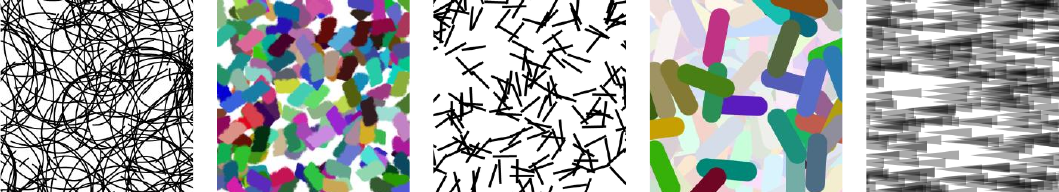}
\caption{Example stroke patches.}
\label{fig:patches}
\end{center}
\end{figure}

\begin{figure}[th]
\begin{center}
\includegraphics[width=0.9\textwidth]{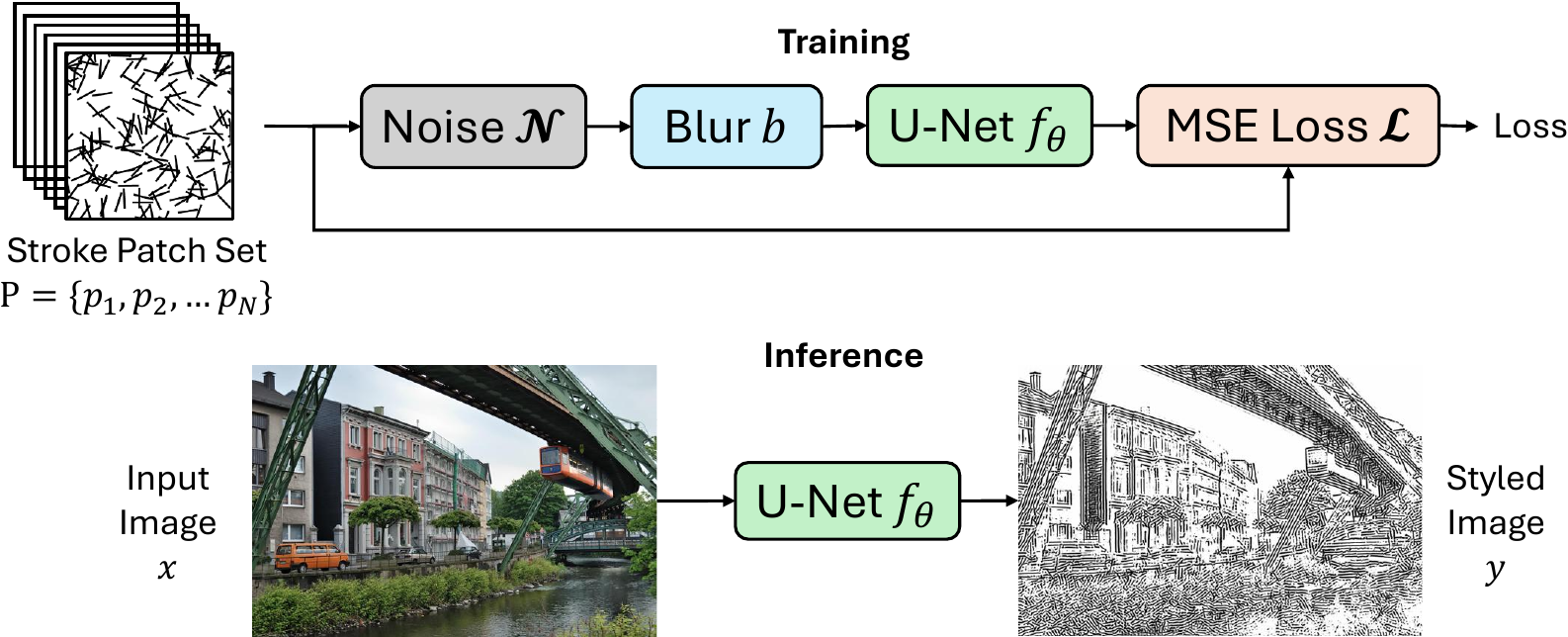}
\caption{\textbf{Training (top)} The training set is a set of stroke patches $P$. Each stroke patch $p_n$ has noise $\mathcal{N}$ added and then is blurred with a Gaussian kernel $b$ and input to U-Net $f_\theta$. The output of these steps is then used to compute the MSE loss with respect to $p_n$. \textbf{Inference (bottom)} An arbitrary input image $x$ is passed through the trained U-Net $f_\theta$ yielding a stylized image $y$.}
\label{fig:system}
\end{center}
\end{figure}
\paragraph{Stroke Patches}
\cref{fig:patches} shows a single stroke patch extracted from five different stroke patch sets.
We create a set of related stroke patch images procedurally by randomly varying various patch parameters such as stroke shape, size, location, orientation, color, and noise level. An algorithm for generating a set of stroke patches for the Wet Brush style is specified in \cref{alg:stroke_patch}.
\paragraph{Architecture}
Our approach is depicted in \cref{fig:system}. The main component is a U-Net \citep{ronneberger2015u} $f$ with parameters $\theta$.
We make two minor changes to the standard U-Net configuration.
The first is that we place a sigmoid operator at the output to ensure that the output values always lie in the 0 to 1 range.
The second is that we have replaced all batch normalization operators \citep{ioffe2015batch} with instance norm operators \citep{ulyanov2016instance} allowing us to train on relatively large stroke patches with small batch sizes on modest GPUs.
During training, we sometimes add noise $\mathcal{N}$ (usually Gaussian or uniform) to the patches and always use a Gaussian blur operator $b$ (typically with radius 5.0) to soften the stroke patch images. 
\paragraph{Training and Inference}
To train $f_\theta$, we minimize the following mean-squared error (MSE) loss:
\begin{equation}
\mathcal{L}(\theta) =\frac{1}{N}\sum_{i=1}^N(f_\theta(b(p_i+\mathcal{N})) - p_i)^2
\end{equation}
At inference, to create an artistic image $y$, we pass an input image $x$ through the trained $f_\theta$ as follows:
\begin{equation}
    y = f_\theta(x).
\end{equation}
During training, the model $f_\theta$ learns a mapping from the blurred and noised stroke patches to the unprocessed stroke patches.
During inference, the model transforms continuous tone areas in the input image that are somewhat similar to a stroke in a stroke patch into a clean, well-defined stroke in the stylized output.
\input{algorithms/stroke_patch_algorithm}
\section{Results}
\label{sec:results}
All training and inference was performed on a Nvidia 4090 GPU.
Unless otherwise stated, we train with a set $P$ of $N$=5000 stroke patches of size 400 $\times$ 400 pixels for 10 epochs using the Adam optimizer \citep{kingma2015adam} with a learning rate of 0.001, and a Gaussian blur ($b$) radius of 5.0.
Training takes 98 seconds per epoch, and inference takes less than a second.
Source code is available at \url{https://github.com/jfb54/stroke_patches}.

\begin{figure}
\begin{center}
\includegraphics[width=1.0\textwidth]{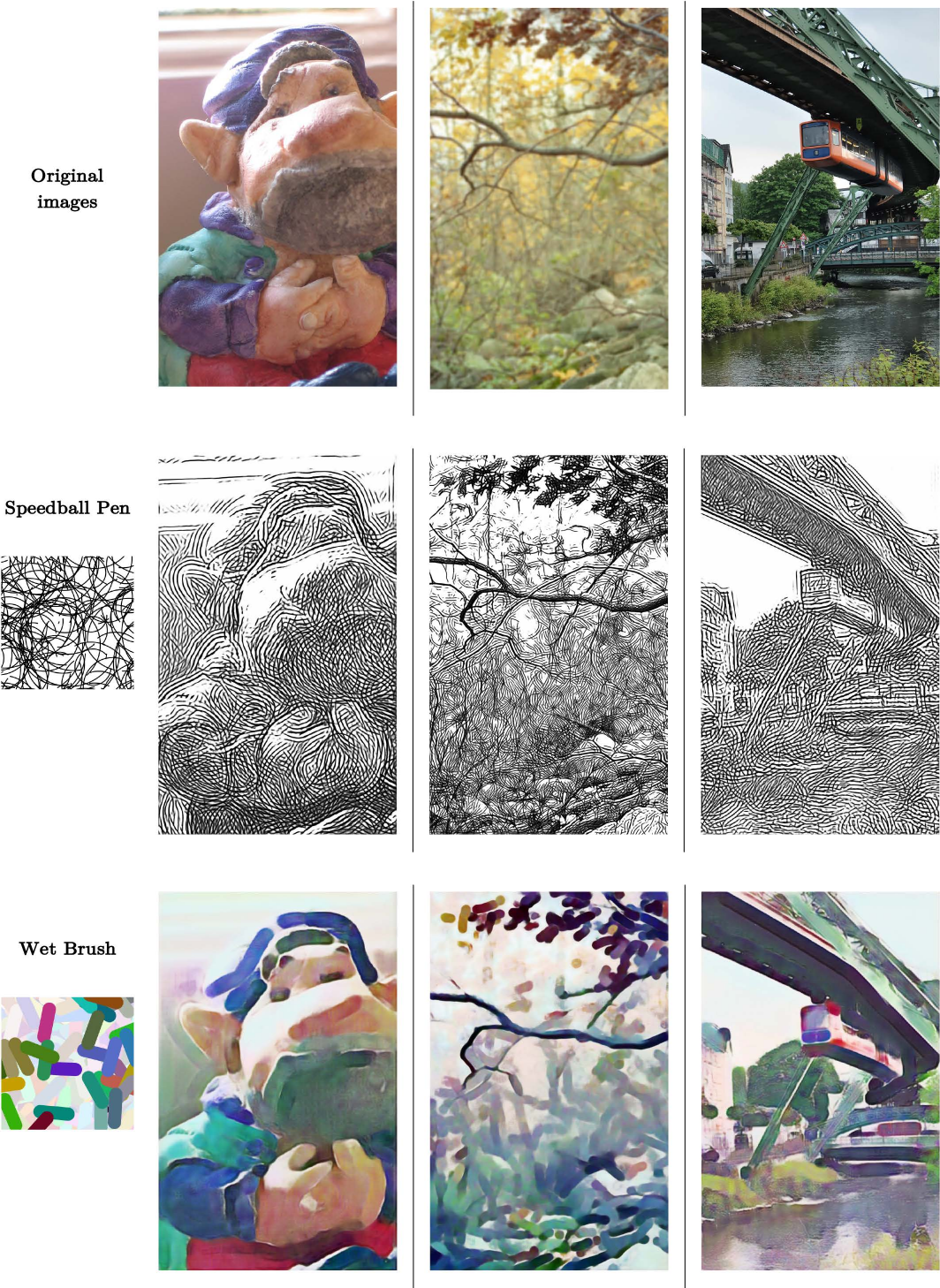}
\caption{\textbf{Results Part 1.} The top row depicts three original unprocessed images and the middle and bottom rows show the results of applying the model trained on two different stroke patches sets (Speedball Pen and Wet Brush). A sample stroke patch is shown to the left of the stylized images. \textbf{Original photo credits:} Ian Jaffray (left, center),  Mbdortmund, GFDL 1.2 \url{http://www.gnu.org/licenses/old-licenses/fdl-1.2.html}, via Wikimedia Commons \url{https://commons.wikimedia.org/wiki/File:Wuppertal-100508-12825-Uferstra\%C3\%9Fe.jpg} (right).}
\label{fig:results_1}
\end{center}
\end{figure}

\begin{figure}
\begin{center}
\includegraphics[width=1.0\textwidth]{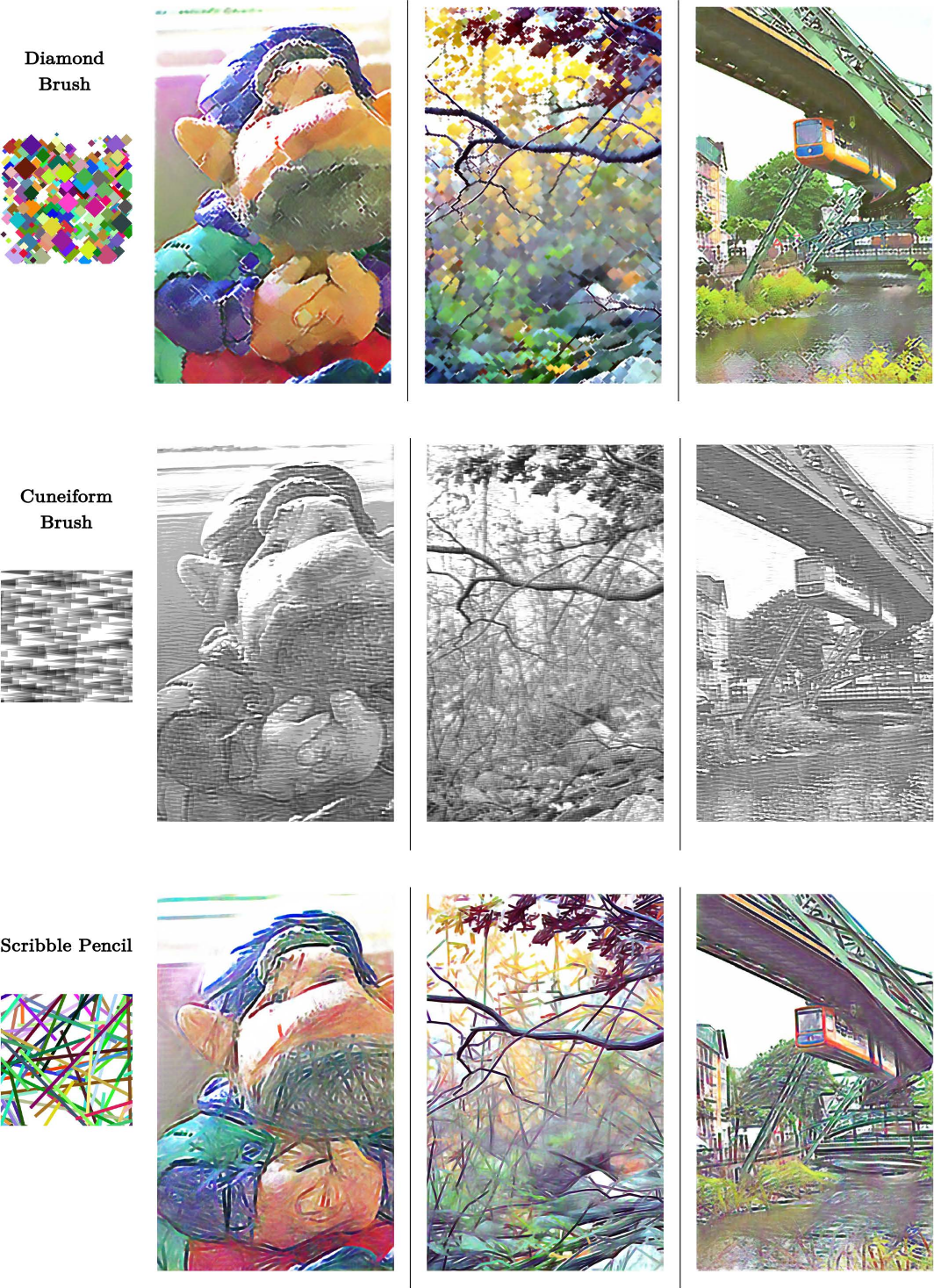}
\caption{\textbf{Results Part 2.} Each row shows the results of applying the model trained on three different stroke patches sets (Diamond Brush, Cuneiform Brush, and Scribble Pencil) using the same original images as \cref{fig:results_1}. A sample stroke patch is shown to the left of the stylized images.}
\label{fig:results_2}
\end{center}
\end{figure}

\begin{figure}
\begin{center}
\includegraphics[width=0.86\textwidth]{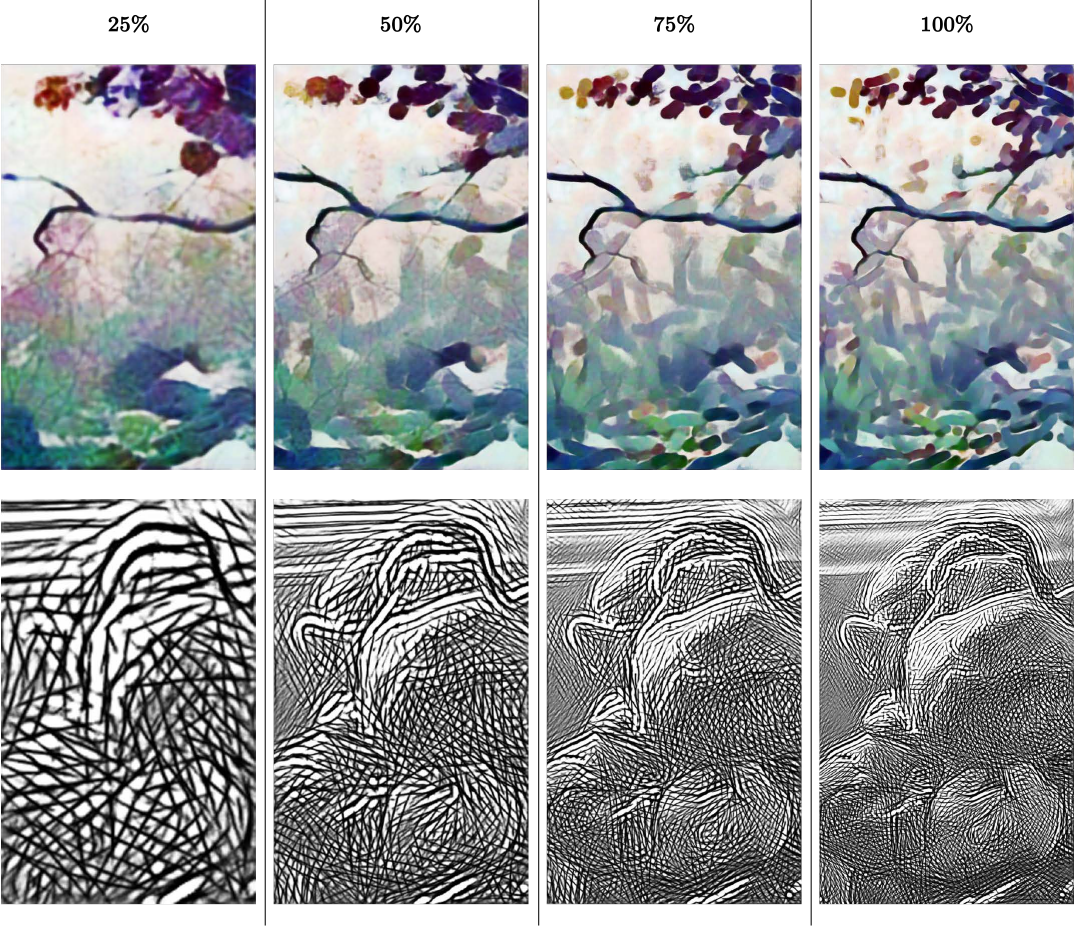}
\caption{\textbf{Effect of scaling the original images.} Each column shows the effect of reducing the original image size by the percentage specified in the column heading before applying the model trained on the Wet Brush (top row) and Speedball Pen (bottom row) stroke patch sets and then enlarging the resulting image back to its original size.}
\label{fig:results_parameters}
\end{center}
\end{figure}

\begin{figure}
\begin{center}
\includegraphics[width=0.86\textwidth]{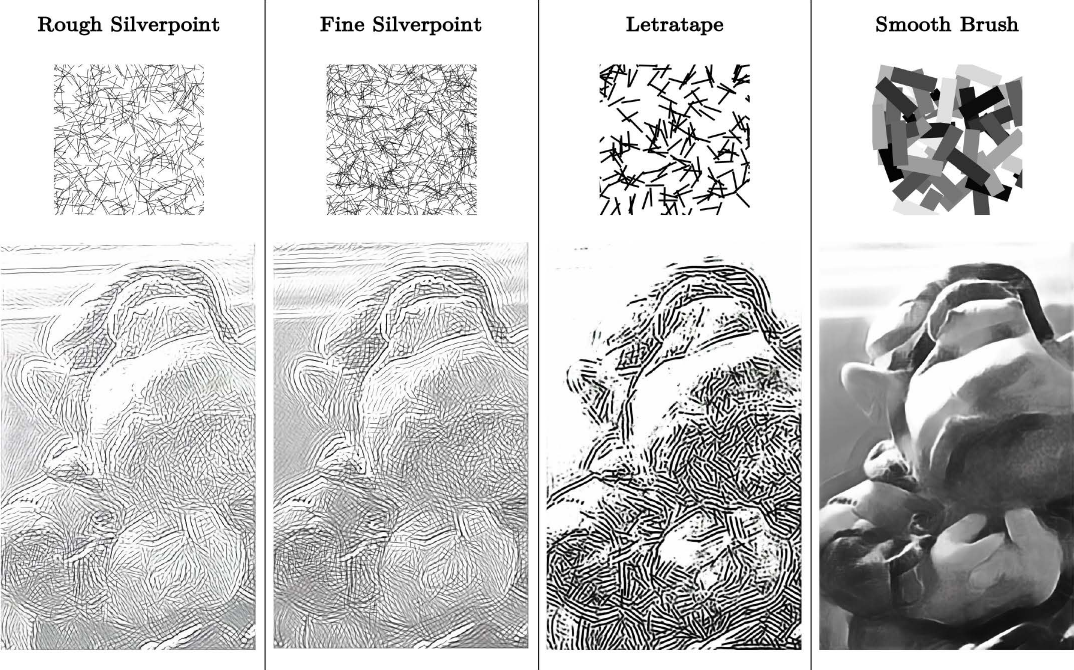}
\caption{\textbf{Effect of changing stroke patch set generation parameters.} The Rough Silverpoint style uses 700 black strokes that are 50 pixels long and 1 pixel wide. The Fine Silverpoint style increases the number of stokes to 1200. The Letratape style decreases the number of strokes to 200, but increases their width to 5 pixels. Finally, the Smooth Brush style reduces the number of strokes to 50, but increases the stroke length to 120 pixels and the stroke width to 40 pixels and adds Gaussian noise with $\mu$=0 and $\sigma$=500.}
\label{fig:results_scaling}
\end{center}
\end{figure}

\cref{fig:results_1,fig:results_2} show three original unprocessed images and stylized results when trained on five different sets of stroke patches.
Despite the simplicity of the procedurally generated stroke patches, the method yields a wide variety of distinctive artistic effects.
\cref{fig:results_scaling} shows how stylized results can be made more prominent by reducing the size of an original image by a fixed fraction $r$, applying the model, and then enlarging the resulting image back to its original size.
More precisely, the output is $y = R_{1/r}(f_\theta(R_r(x)))$, where $R_r$ indicates the spatial image scaling operator by a factor $0 < r \le 1$.
\cref{fig:results_parameters} shows the effect of modifying various stroke patch set generation parameters including the number of the strokes in each patch, the stroke width and length, the amount of Gaussian noise added, and the stroke opacity.
The specification of these parameters allows for a significant degree of control over the result.
\section{Summary, Limitations, and Societal Impact}
\label{sec:discussion}
\textbf{Summary }In this work, we presented a regression-based method to artistically render an image that allows for explicit control over the stroke composition and level of detail in the output through the use of an extensible set of stroke patches.

\textbf{Limitations} The primary limitation of this work is that, unlike style transfer or diffusion methods, the overall high-level look of the image cannot be specified by a text prompt or reference to another image.
Our work requires the design and specification of low-level detail via a set of stroke patches.
Our approach does not replace existing style transfer, diffusion-based, or other effect methods, but instead complements them, giving the digital artist a new customizable tool with a high degree of control to transform images or video to have an artistic or painterly appearance.

\textbf{Societal Impact} We hope that our work will be useful for both amateur and professional artists to create artistic imagery, perhaps saving time and money when compared to hiring an artist who works in traditional media.
We see no negative societal impact as our model is limited in capability and it would not be possible to use the model to generate counterfeit works.
\clearpage
\newpage

\bibliography{references}
\bibliographystyle{unsrtnat}





\newpage
\section*{NeurIPS Paper Checklist}
\begin{enumerate}

\item {\bf Claims}
    \item[] Question: Do the main claims made in the abstract and introduction accurately reflect the paper's contributions and scope?
    \item[] Answer: \answerYes{} 
    \item[] Justification: We claim three contributions in our paper in \cref{sec:introduction} and each is described in detail in \cref{sec:method}.

\item {\bf Limitations}
    \item[] Question: Does the paper discuss the limitations of the work performed by the authors?
    \item[] Answer: \answerYes{} 
    \item[] Justification: Refer to \cref{sec:discussion} where we discuss limitations of our work.

\item {\bf Theory assumptions and proofs}
    \item[] Question: For each theoretical result, does the paper provide the full set of assumptions and a complete (and correct) proof?
    \item[] Answer: \answerNA{} 
    \item[] Justification: Our paper does not include theoretical results.

    \item {\bf Experimental result reproducibility}
    \item[] Question: Does the paper fully disclose all the information needed to reproduce the main experimental results of the paper to the extent that it affects the main claims and/or conclusions of the paper (regardless of whether the code and data are provided or not)?
    \item[] Answer: \answerYes{} 
    \item[] Justification: We supply all experimental setting/details required to reproduce the experiments. We do this by supplying full source code as well as thorough information in \cref{sec:method,sec:results,alg:stroke_patch}.

\item {\bf Open access to data and code}
    \item[] Question: Does the paper provide open access to the data and code, with sufficient instructions to faithfully reproduce the main experimental results, as described in supplemental material?
    \item[] Answer: \answerYes{} 
    \item[] Justification: We provide source code to reproduce the results at \url{https://github.com/jfb54/stroke_patches}. Along with the code, we provide a README file that details installation, configuration, and options in order to reproduce the results.

\item {\bf Experimental setting/details}
    \item[] Question: Does the paper specify all the training and test details (e.g., data splits, hyperparameters, how they were chosen, type of optimizer, etc.) necessary to understand the results?
    \item[] Answer: \answerYes{} 
    \item[] Justification: We supply all experimental setting/details required to reproduce the results. We do this by supplying full source code as well as thorough information in \cref{sec:method,sec:results}.

\item {\bf Experiment statistical significance}
    \item[] Question: Does the paper report error bars suitably and correctly defined or other appropriate information about the statistical significance of the experiments?
    \item[] Answer: \answerNo{} 
    \item[] Justification: Our experimental results contain only image outputs. No numerical results are stated. 

\item {\bf Experiments compute resources}
    \item[] Question: For each experiment, does the paper provide sufficient information on the computer resources (type of compute workers, memory, time of execution) needed to reproduce the experiments?
    \item[] Answer: \answerYes{} 
    \item[] Justification: \cref{sec:results} details compute resources used and processing times needed to reproduce the experiments. The full research project required more compute than the experiments reported in the paper due to early and failed experiments that didn't make it into the paper.
    
\item {\bf Code of ethics}
    \item[] Question: Does the research conducted in the paper conform, in every respect, with the NeurIPS Code of Ethics \url{https://neurips.cc/public/EthicsGuidelines}?
    \item[] Answer: \answerYes{} 
    \item[] Justification: The authors have reviewed the NeurIPS Code of Ethics and have complied with them.

\item {\bf Broader impacts}
    \item[] Question: Does the paper discuss both potential positive societal impacts and negative societal impacts of the work performed?
    \item[] Answer: \answerYes{} 
    \item[] Justification: Refer to \cref{sec:discussion} for a discussion of positive and negative societal impact of our work.
    
\item {\bf Safeguards}
    \item[] Question: Does the paper describe safeguards that have been put in place for responsible release of data or models that have a high risk for misuse (e.g., pretrained language models, image generators, or scraped datasets)?
    \item[] Answer: \answerNA{} 
    \item[] Justification: The models that we will release are low-risk.

\item {\bf Licenses for existing assets}
    \item[] Question: Are the creators or original owners of assets (e.g., code, data, models), used in the paper, properly credited and are the license and terms of use explicitly mentioned and properly respected?
    \item[] Answer: \answerYes{} 
    \item[] Justification: In our paper we use the following assets:
        \begin{itemize}
            \item We properly credit the original photos that we use for styling in a caption where they are first used.
            \item We properly credit and respect the license for the open-source code to implemented the U-Net in our source code.
        \end{itemize}

\item {\bf New assets}
    \item[] Question: Are new assets introduced in the paper well documented and is the documentation provided alongside the assets?
    \item[] Answer: \answerYes{} 
    \item[] Justification: We introduce the following assets, which are fully documented and licensed in our source code repository:
    \begin{itemize}
        \item Source code for reproducing the results.
        \item Pretrained models for the styles used in the paper.
    \end{itemize}

\item {\bf Crowdsourcing and research with human subjects}
    \item[] Question: For crowdsourcing experiments and research with human subjects, does the paper include the full text of instructions given to participants and screenshots, if applicable, as well as details about compensation (if any)? 
    \item[] Answer: \answerNA{} 
    \item[] Justification:Our paper does not involve crowdsourcing nor research with human subjects.

\item {\bf Institutional review board (IRB) approvals or equivalent for research with human subjects}
    \item[] Question: Does the paper describe potential risks incurred by study participants, whether such risks were disclosed to the subjects, and whether Institutional Review Board (IRB) approvals (or an equivalent approval/review based on the requirements of your country or institution) were obtained?
    \item[] Answer: \answerNA{} 
    \item[] Justification: Our paper does not involve crowdsourcing nor research with human subjects.

\item {\bf Declaration of LLM usage}
    \item[] Question: Does the paper describe the usage of LLMs if it is an important, original, or non-standard component of the core methods in this research? Note that if the LLM is used only for writing, editing, or formatting purposes and does not impact the core methodology, scientific rigorousness, or originality of the research, declaration is not required.
    \item[] Answer: \answerNA{} 
    \item[] Justification: This work did not involve any use of LLMs.

\end{enumerate}

\end{document}

%% file: algorithms/stroke_patch_algorithm.tex
\begin{algorithm*}[t]
    \caption {Wet Brush stroke patch set generation}
    \label{alg:stroke_patch}
    \begin{algorithmic}[1]
    \Require $W, H$: stroke patch width and height in pixels (default = 400)
    \Require $N$: number of stroke patches in stroke patch set (default = 5000)
    \Require $BG$: background color (default = 'white'); $L$: stroke length (default = 80)
    \Require $S$: number of strokes in each patch (default = 50); $T$: stroke thickness (default = 40)
    \Procedure{CreateWetBrushStrokePathSet}{$W, H, N, BG, S, T$}
        \State stroke\_patch\_set $\gets$ [\ ] \Comment{set of stroke patches}
        \For{$i \gets 1$ to $N$} \Comment{create N patches}
            \State patch $\gets$ \texttt{create\_image}($W, H, BG$)
            \For{$i \gets 1$ to $S$} \Comment{create S strokes in each patch}
                \State $\phi \gets \mathcal{U}(0,2\pi)$ \Comment{line orientation}
                \State $x_1 \gets \mathcal{U}(-L/2,W+L/2), y_1 \gets \mathcal{U}(-L/2,H+L/2)$ \Comment{line start}
                \State $x_2 = x_1 + L\cos(\phi), y_2 = y_1 + L\sin(\phi)$ \Comment{line end}
                \State $C$ $\gets$ \texttt{SetRGBAColor}$(\mathcal{U}(0,1), \mathcal{U}(0,1), \mathcal{U}(0,1),1)$ \Comment{Random opaque RGB color.}
                \State patch.\texttt{DrawLine}($x_1, y_1, x_2, y_2, L, T, C,$ \texttt{round\_end\_cap})
            \EndFor
            \State stroke\_patch\_set.\texttt{append}(patch)
        \EndFor
        \State \texttt{return}(stroke\_patch\_set)
    \EndProcedure
    \State $\mathcal{N}$ is Gaussian with $\mu = 0$ and $\sigma=500$. Gaussian blur $b$ radius is set to 5.0.
    \end{algorithmic}
\end{algorithm*}